\documentclass{PoS}

\title{Quarkonium production in proton-proton collisions with ALICE at the LHC}

\ShortTitle{ALICE quarkonia in pp}

\author{\speaker{Philippe Rosnet}, on behalf of the ALICE Collaboration, \\
        Laboratoire de Physique de Clermont (LPC), Universit\'e Clermont Auvergne, CNRS/IN2P3, Clermont-Ferrand, France \\
        E-mail: \email{rosnet@in2p3.fr}}

\abstract{
ALICE at the LHC has a unique potential to study proton-proton collisions with the goal to probe Quantum ChromoDynamics (QCD). 
The apparatus was designed to reconstruct particles over a large range in transverse momentum and rapidity. 
In particular, quarkonia are very interesting probes of QCD, because their production mechanisms are governed by both perturbative and non-perturbative QCD processes. 
In ALICE, quarkonia are reconstructed via their dilepton decay channel down to zero transverse momentum. 
This contribution gives a short overview of quarkonium production results in proton-proton collisions with ALICE and a comparison to other experimental results and to theoretical models.
}

\FullConference{XXV International Workshop on Deep-Inelastic Scattering and Related Subjects\\
		3-7 April 2017\\
		University of Birmingham, UK}

\begin{document}

\section{Introduction}

Hadronic colliders, like the Large Hadron Collider (LHC) at CERN, are well-suited tools to study Quantum ChromoDynamics (QCD), the theory of the strong interaction, one of the two pillars of the Standard Model of particle physics. 
In particular, quarkonia, which are bound states of heavy quarks Q$\overline{\rm Q}$, i.e. charm (c) or bottom (b), involve in their production mechanism both perturbative (pQCD) and non-perturbative aspects of QCD.
Due to their large mass ($m_{\rm c} \approx 1.3$~GeV$/c^2$ and $m_{\rm b} \approx 4.4$~GeV$/c^2$), heavy quarks are produced in initial hard processes (pQCD), mainly via gluon fusion ${\rm g}\,{\rm g} \to {\rm Q}\,\overline{\rm Q} + X$ at the LHC~\cite{Mangano1997}, while their hadronisation in quarkonium is a soft process described by non-perturbative approaches.

For 40 years, different theoretical models have been developped as an attempt to describe the whole production mechanism from partonic interaction to heavy quark hadronisation in quarkonium. 
All approaches assume factorisation between hard and soft scales.
The Color-Evaporation Model (CEM)~\cite{Fritzch1977} is based on the quark-hadron duality where each Q$\overline{\rm Q}$ pair evolves in a bound state below the open heavy-flavour threshold with a phenomenological probability factor (energy and process independent) for each quarkonium state. 
The Color-Singlet Model (CSM)~\cite{EinhornEllis1981} assumes no evolution of the quantum color-singlet state between the Q$\overline{\rm Q}$ production and the quarkonium formation with a wave function computed at zero Q$\overline{\rm Q}$ separtion, i.e. without any free parameter. 
The Color-Octet Mechanism (COM)~\cite{BodwinBraatenLepage1995} is based on a Non Relativistic QCD (NRQCD) approach introducing Long-Distance Matrix Elements (LDMEs) for the hadronisation probability in a quarkonium state; the LDMEs are determined from experimental data.
One of the most relevant observables to discriminate between the different theoretical models is the polarisation state which is predicted to be strongly transverse-momentum dependent for S-wave states -- J/$\psi$, $\psi$(2S) and $\Upsilon$(nS) -- with two opposite trends: partially longitudinal in CSM and partially transverse in COM.

Beyond the fact that quarkonia represent a QCD laboratory, their understanding in basic hadronic collisions, i.e. in proton-proton (pp) collisions at the LHC, is of crucial importance for the interpretation of proton-nucleus (pA) and heavy-ion collisions (AA) in view of the study of medium effects~\cite{SaporeGravis2016}: "cold nuclear matter" in pA and "hot and dense nuclear matter" (like the Quark-Gluon Plasma) in AA collisions.

\section{Experimental conditions}

The ALICE detector~\cite{ALICE_detector} is particularly well suited for the reconstruction of low transverse momentum particles and in this way to measure the bulk of the production cross section of hadrons.
Quarkonia are reconstructed in their dielectron channel at mid-rapidity ($|y| < 0.9$) with the Inner Tracking System (ITS) and the Time Projection Chamber (TPC), and in their dimuon channel at forward rapidity ($2.5 < y < 4.0$) with the muon spectrometer.
The minimum-bias trigger uses the mid-rapidity Silicon Pixel Detector (SPD) and the forward/backward V0 hodoscopes for the dielectron channel. 
This trigger in association with a dedicated dimuon trigger is used to increase the statistics in the dimuon channel. 
The SPD is also used to separate prompt J/$\psi$ and those from b-hadron decay at mid-rapidity, as illustrated in Fig.~\ref{fig1} (left).
The higher statistics recorded at forward rapidity allows us to study the full dimuon resonance spectrum from $\omega$ to $\Upsilon$(nS), as shown in Fig.~\ref{fig2} (left).

ALICE has recorded LHC proton-proton collisions at centre-of-mass energies $\sqrt{s} = 0.9$, 2.76, 5.02, 7, 8 and 13~TeV, with minimum-bias integrated luminosity $L_{\rm int}$ varying from about 1~nb$^{-1}$ to 3~pb$^{-1}$.

\section{Cross section measurements}

Quarkonium cross sections $\sigma_{\cal Q}$ are measured in a differential way as a function of rapidity ($y$) and transverse momentum ($p_{\rm T}$) from the number of reconstructed quarkonia $N_{\cal Q}$ corrected by the acceptance times efficiency $A\cdot\epsilon$ and the associated dilepton branching ratio BR$_{{\cal Q}\to l^+l^-}$:
\begin{eqnarray*}
\frac{{\rm d}^2\sigma_{\cal Q}}{{\rm d}p_{\rm T}{\rm d}y} = \frac{1}{\Delta y \Delta p_{\rm T}} \frac{1}{L_{\rm int}} \frac{N_{\cal Q}(y,p_{\rm T})}{{\rm BR}_{{\cal Q}\to l^+l^-} A\cdot\epsilon(y,p_{\rm T})} 
\end{eqnarray*} 
where $\Delta y$ and $\Delta p_{\rm T}$ are the widths of the rapidity and transverse momentum bins, respectively.
The accep\-tan\-ce-efficiency correction factor is determined from realistic run-by-run simulations to account for the variations in the detector configuration during the acquisition period.

\begin{figure}[h]
\centering
\includegraphics[height=6cm,clip]{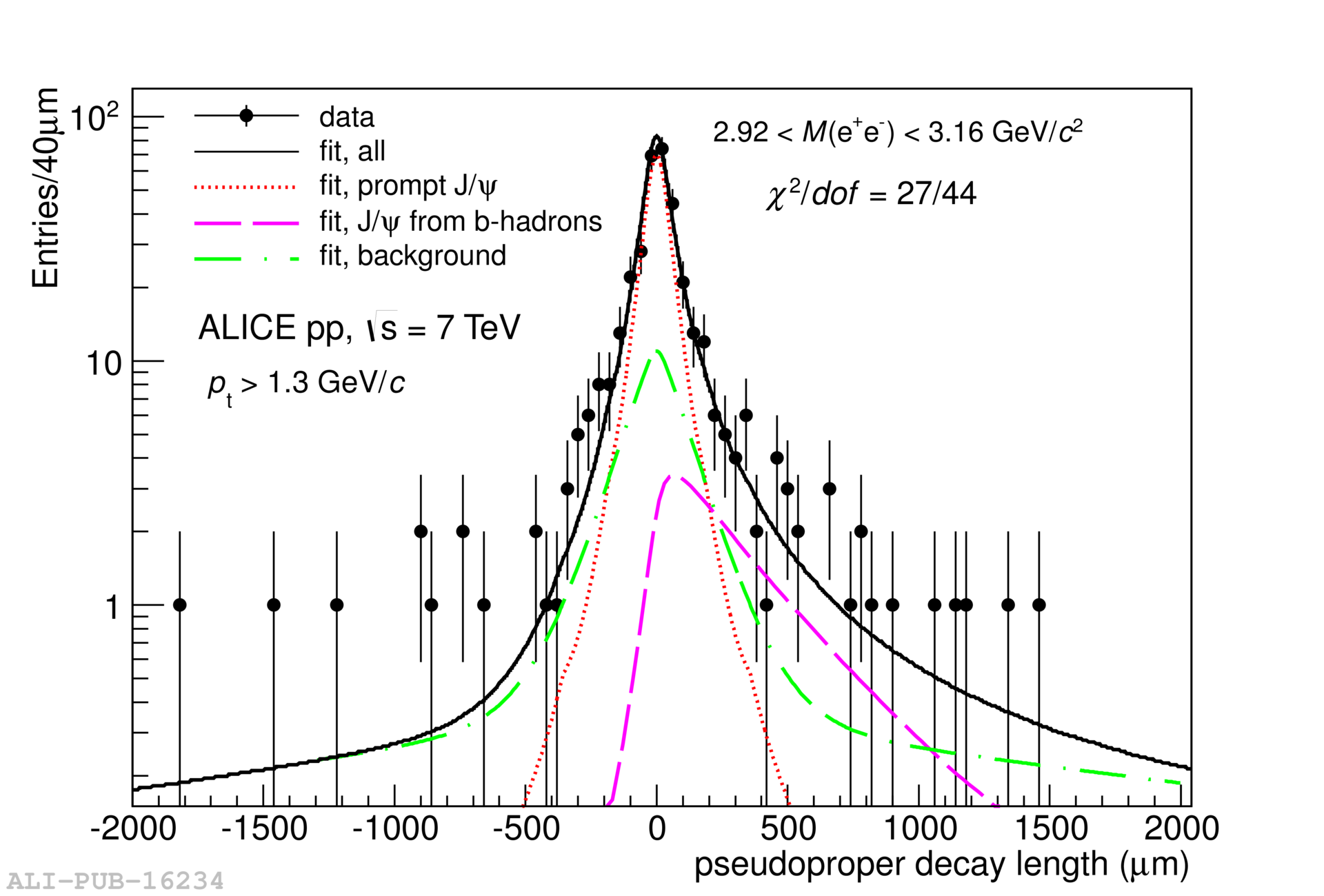}
\hfill
\includegraphics[height=6cm,clip]{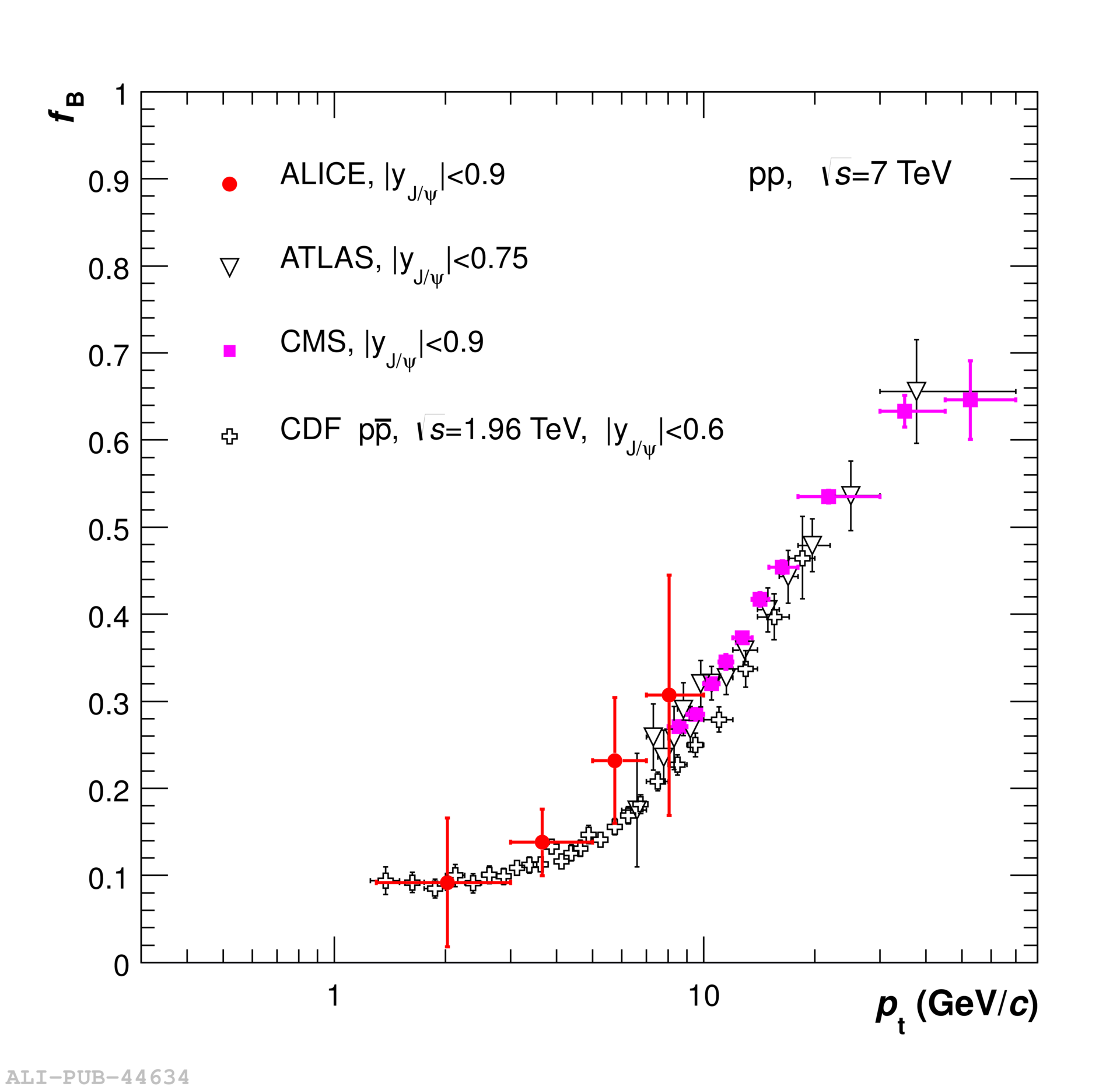}
\caption{Left~\cite{ALICEprompt2012}: dielectron pseudoproper decay length, a kinematical variable based on the distance between the primary and displaced vertices, allowing us to separate prompt J/$\psi$ and those from b-hadron decay. Right~\cite{ALICEprompt2012}: fraction $f_{\rm B}$ of J/$\psi$ from b-hadron decay at mid-rapidity ($|y| < 0.9$) as a function of $p_{\rm T}$ at $\sqrt{s} = 7$~TeV compared to other experimental results.}
\label{fig1}    
\end{figure}

Figure~\ref{fig1} (right) shows the fraction $f_{\rm B}$ of J/$\psi$ from b-hadron decay at mid-rapidity as a function of transverse momentum. 
The ALICE data points are compared to those from other LHC experiments and illustrate the complementarity to cover the low-$p_{\rm T}$ range $1.3 < p_{\rm T} < 10$~GeV/$c$ where a good agreement with ATLAS and CMS results is observed.
Moreover, results from ALICE are in good agreement with those from CDF (p$\bar{\rm p}$ experiment at lower energy).
All results exhibit a clear increase of $f_{\rm B}$ as a function of $p_{\rm T}$ which seems to be proportional to $\ln(\sqrt{s})$ for $p_{\rm T} > 10$~GeV/$c$.

\begin{figure}[h]
\centering
\includegraphics[height=5cm,clip]{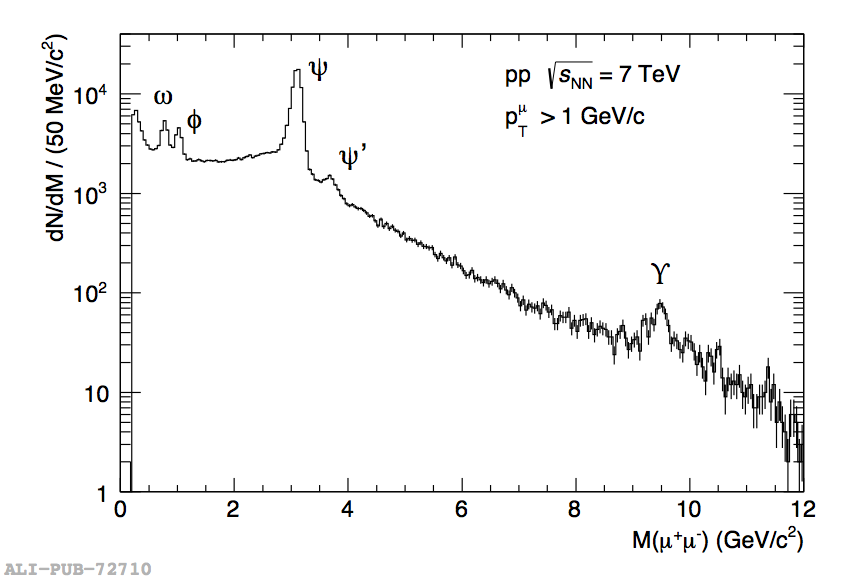}
\hfill
\includegraphics[height=5cm,clip]{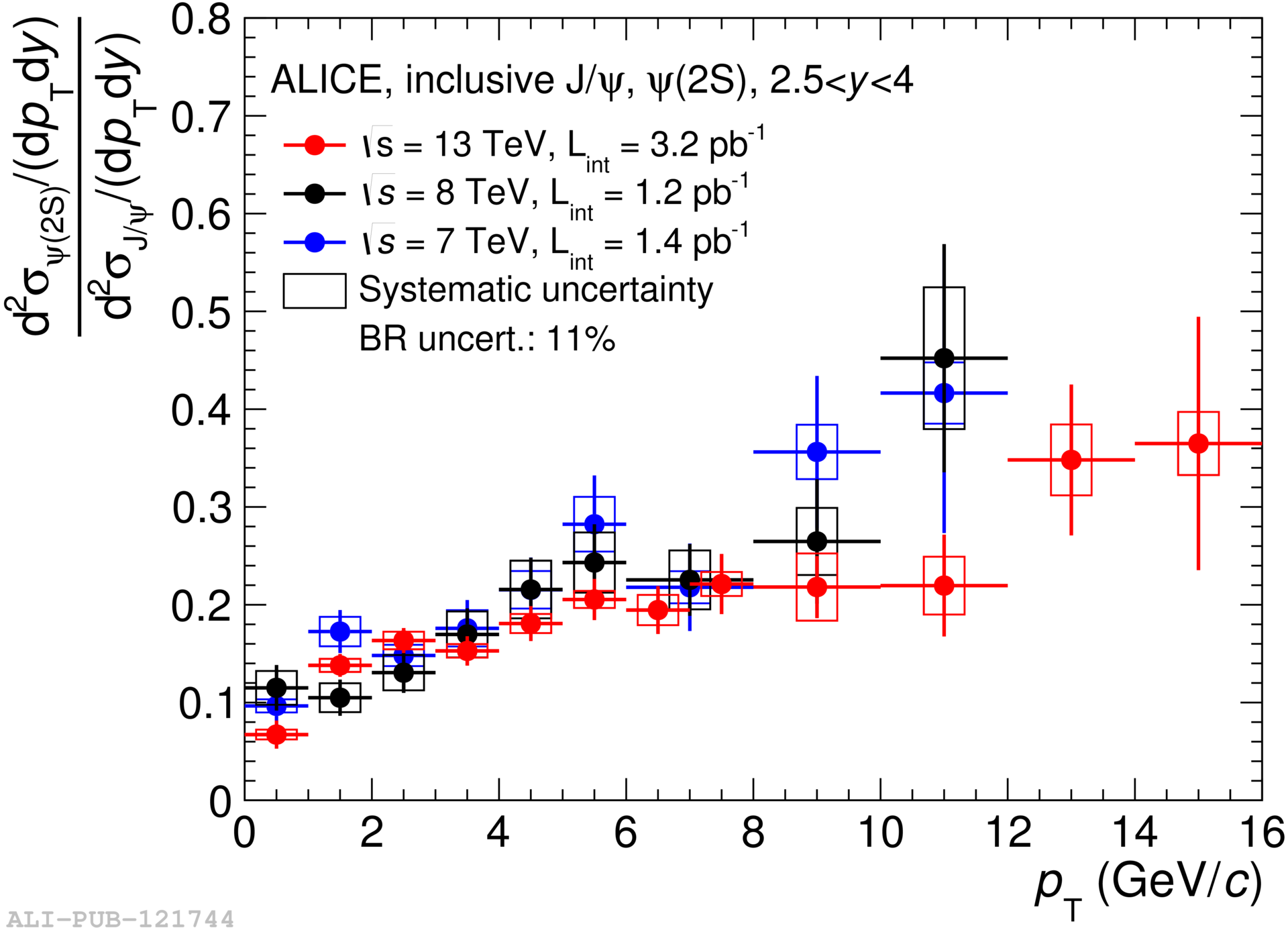}
\caption{Left~\cite{ALICE_performances}: dimuon spectrum at forward rapidity $2.5 < y < 4.0$ for $\sqrt{s} = 7$~TeV.  Right~\cite{ALICEpp13TeV}: ratio of $\psi$(2S) and J/$\psi$ $p_{\rm T}$-differential cross section at forward rapidity and for different centre-of-mass energies.}
\label{fig2}    
\end{figure}

For what concerns charmonium excited states, forward rapidity measurements show that the $\psi$(2S) production cross section is about one order of magnitude lower than the inclusive J/$\psi$ one, as illustrated in Fig.~\ref{fig2} (right) for three LHC energies, but with an increase of the $\psi$(2S) fraction with transverse momentum.
This behaviour seems to be independent of the energy and is explained by the non negligible fraction of prompt J/$\psi$ coming from the decay of high-mass c$\bar{\rm c}$ resonances -- about 16\% from $\chi_{\rm c}$(1P)~\cite{LHCbChic} and 8\% from $\psi$(2S)~\cite{LHCbPsi2S} -- which soften the $p_{\rm T}$ spectrum of inclusive J/$\psi$.
The compilation of inclusive J/$\psi$ results at forward rapidity for different energies $\sqrt{s} = 2.76$~\cite{ALICEpp2TeV}, 5.02~\cite{ALICEpp13TeV}, 7~\cite{ALICEpp7TeV}, 8~\cite{ALICEpp8TeV}, 13~TeV~\cite{ALICEpp13TeV} shows that the $p_{\rm T}$ trend exhibits a change of exponential slope for $p_{\rm T} \sim 10$~GeV/$c$.
This observation is attributed to the non-prompt J/$\psi$ contribution characterised by a harder $p_{\rm T}$ spectrum.
  
\begin{figure}[h]
\centering
\includegraphics[height=5.5cm,clip]{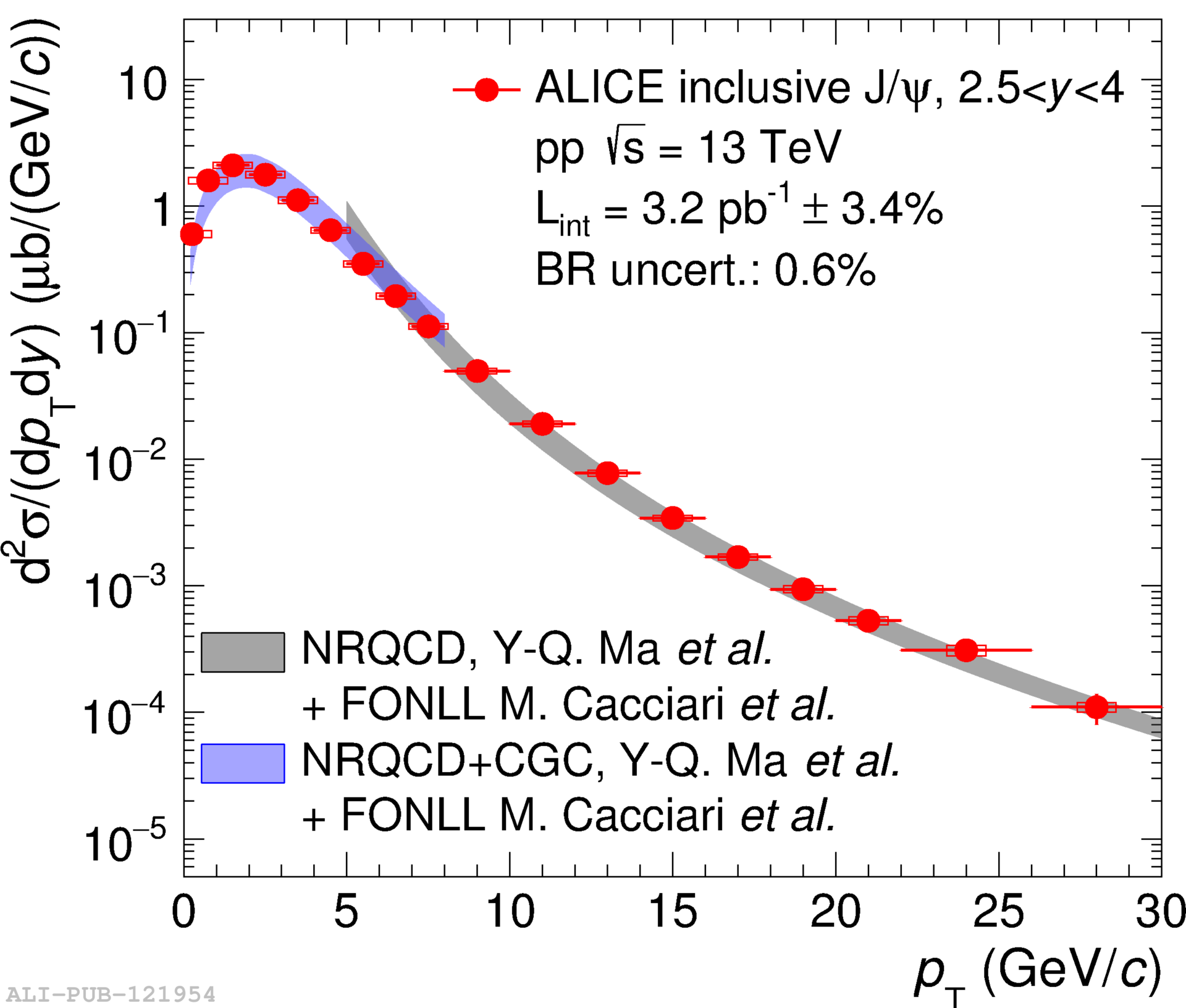}
\hfill
\includegraphics[height=5.5cm,clip]{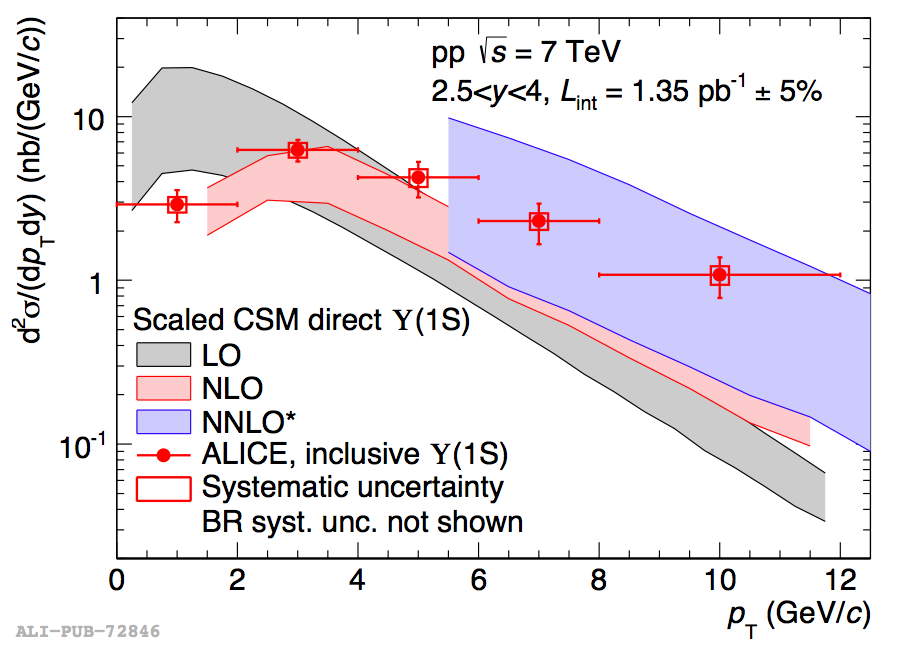}
\caption{Left~\cite{ALICEpp13TeV}: inclusive J/$\psi$ $p_{\rm T}$-differential cross section at forward rapidity at $\sqrt{s} = 13$~TeV compared to an ad-hoc mixed model (see text).  Right~\cite{ALICEpp7TeV}: inclusive $\Upsilon$(1S) $p_{\rm T}$-differential cross section at forward rapidity at $\sqrt{s} = 7$~TeV compared to a CSM prediction (NNLO$^\star$ indicates that only the main contributing diagrams are taken into account at this order of the calculation).}
\label{fig3}    
\end{figure}

The comparison of the $p_{\rm T}$-differential cross section with CSM, and illustrated for $\Upsilon$(1S) in Fig.~\ref{fig3} (right), shows that higher-order QCD processes (NLO and main NNLO contributions) are important.
An ad-hoc model~\cite{MaVenugopalan2014} mixing a COM prediction with a Color Glass Condensate (CGC) approach to describe the low-$p_{\rm T}$ region and the Fixed-Order Next-to-Leading Logarithm (FONLL) prediction to account for the non-prompt component is able to describe the full $p_{\rm T}$-differential cross section of inclusive J/$\psi$, as illustrated in Fig.~\ref{fig3} (left). 
On the other hand, the $p_{\rm T}$-integrated cross section in the rapidity range $2.5 < y < 4.0$ shows that the CEM underestimates the total cross section at forward rapidity for the highest LHC energy~\cite{ALICEpp13TeV}. 

\section{Polarisation study}

\begin{figure}[h]
\centering
\includegraphics[height=5cm,clip]{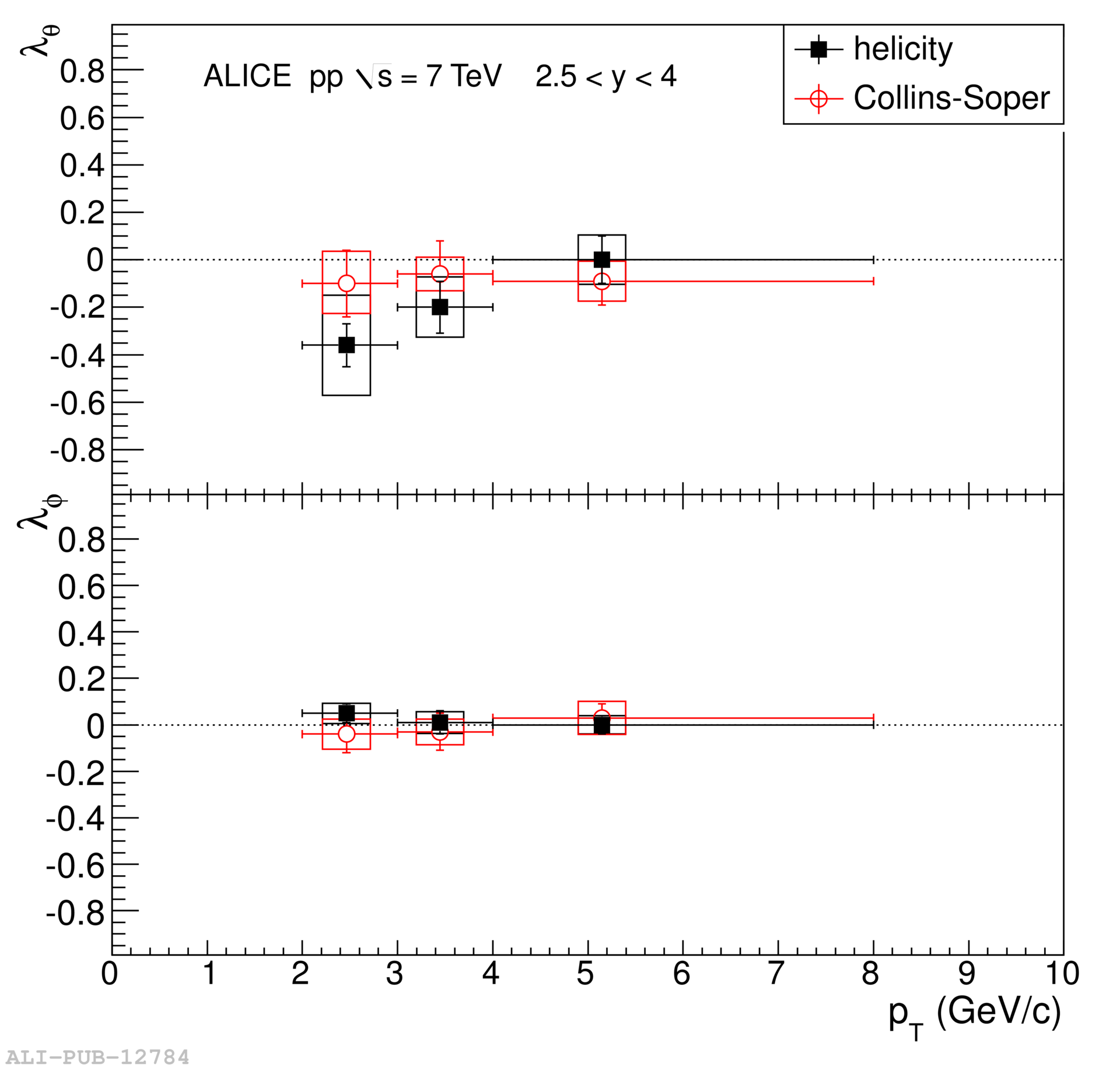}
\hfill
\includegraphics[height=5cm,clip]{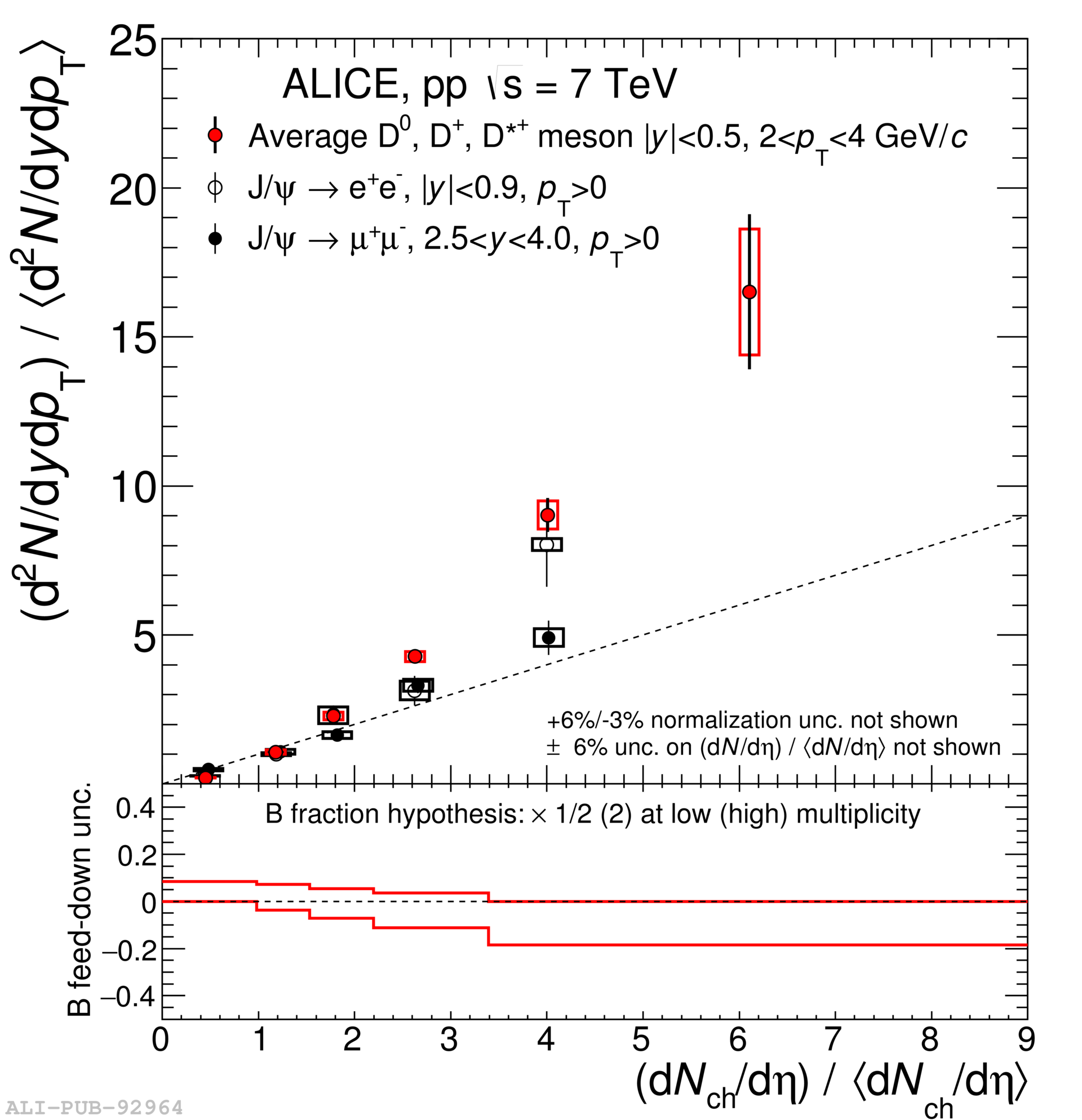}
\caption{Left~\cite{ALICEpolarization7TeV}: Main ($\lambda_\theta$) and second (here noted $\lambda_\phi$) J/$\psi$ polarisation parameters as a function of $p_{\rm T}$ at forward rapidity and $\sqrt{s} = 7$~TeV in both, the helicity and the Collins-Soper frames.  Right~\cite{ALICEmpi7TeV}: relative J/$\psi$ (and D-meson) yield as a function of the event relative charged particle multiplicity $N_{\rm ch}$.}
\label{fig4}    
\end{figure}

The study of polarisation of S-wave states (spin 1) is carried out by the analysis of the dilepton decay distribution in the quarkonium rest frame using spherical coordinates $(\theta,\varphi)$ with respect to different reference axes (e.g. helicity or Collins-Soper frames): $W(\theta,\varphi) \propto \frac{1}{3+\lambda_\theta} [1 + \lambda_\theta\cos^2\theta + \lambda_\varphi\sin^2\theta\cos(2\varphi) + \lambda_{\theta\varphi}\sin(2\theta)\cos\varphi]$, where $\lambda_\theta$, $\lambda_\varphi$ and $\lambda_{\theta\varphi}$ are the main, the second and the third polarisation parameters, respectively.
The question of quarkonium polarisation was already addressed by pre-LHC experiments, especially at Tevatron, but experimental results were not always in agreement~\cite{CDF}.
In the LHC era, the improvement of analysis procedures has allowed us to converge to robust experimental results.
ALICE has performed the first J/$\psi$ polarisation measurement for pp at $\sqrt{s} = 7$~TeV~\cite{ALICEpolarization7TeV} in both, the helicity and the Collins-Soper frames in the transverse momentum range $2 < p_{\rm T} < 8$~GeV/$c$, as illustrated in Fig.~\ref{fig4} (left). 
Results show that no polarisation is observed in pp collisions at this LHC energy, i.e. in contradiction with initial CSM and COM predictions. 
This result is confirmed by the other LHC experiments over a very large momentum range (up to 70~GeV/$c$) and for different S-wave states, as summarised in reference~\cite{Lourenco2014}.

\section{Event multiplicity dependence}

In high-energy pp collisions, Multi-Parton Interactions (MPI)~\cite{Bernhard2009}, i.e. several hard partonic interactions occurring in a single pp collision, are expected to take place due to the high parton densities at low $x$-Bjorken (at the LHC, J/$\psi$ production at $y=4$ involves typically gluons with $x$-Bjorken $\sim 10^{-5}$).
So, a dependence of quarkonium (and other heavy hadron) production is expected with respect to the underlying event.
Experimentally, the underlying event activity is quantified by the event multiplicity. 
Figure~\ref{fig4} (right) shows the relative J/$\psi$ yield, i.e. the J/$\psi$ multiplicity normalised to its mean  multiplicity, as a function of the event relative charged particle multiplicity.
The behaviour of D-mesons is also shown for comparison. 
The observed non-linear increase of the relative J/$\psi$ yield with the relative charged multiplicity is similar to that of the D mesons.
The comparison with Monte Carlo event generators shows that only generators including MPI processes are able to match with this trend.

\section{Conclusions}

LHC experiments have already provided a large set of data over a wide energy range and with different beam species.
In proton-proton collisions, the particularities of the ALICE detector have allowed us to address complementary aspects of QCD compared to the other LHC experiments. 
In the quarkonium sector, the measurements of differential cross sections, quarkonium state cross section ratios and polarisation parameters give numerous inputs to constrain the different theoretical approaches. 
It seems that an ad-hoc mixed picture CGC+COM+FONLL is able to describe the full $p_{\rm T}$ spectrum of high-precision inclusive J/$\psi$ measurements, but the question of quarkonium polarisation remains still open.
The LHC provides new opportunities to study the MPI phenomenon.
Detailed understanding requires higher statistics which is the scope of the future LHC runs coming with upgraded detectors to improve the quality of present measurements and to disentangle processes, e.g. prompt and non-prompt J/$\psi$ at forward rapidity with the future ALICE Muon Forward Tracker~\cite{ALICE-MFT}.

\end{document}